# "Bayesian Marketing Mix Modeling" for Lemonade

Roy Ravid
*Lemonade, Inc*
roy.ravid@lemonade.com

## Abstract

Marketing mix modeling (MMM) is a widely used method to assess the effectiveness of marketing campaigns and optimize marketing strategies. Bayesian MMM is an advanced approach that allows for the incorporation of prior information, uncertainty quantification, and probabilistic predictions [1]. In this paper, we describe the process of building a Bayesian MMM model for the online insurance company Lemonade.

We first collected data on Lemonade's marketing activities, such as online advertising, social media, and brand marketing, as well as performance data. We then used a Bayesian framework to estimate the contribution of each marketing channel on total performance, while accounting for various factors such as seasonality, market trends, and macroeconomic indicators.

To validate the model, we compared its predictions with the actual performance data from A/B-testing and sliding window holdout data [2]. The results showed that the predicted contribution of each marketing channel is aligned with A/B test performance and is actionable.

Furthermore, we conducted several scenario analyses using convex optimization to test the sensitivity of the model to different assumptions and to evaluate the impact of changes in the marketing mix on sales. The insights gained from the model allowed Lemonade to adjust their marketing strategy and allocate their budget more effectively.

Our case study demonstrates the benefits of using Bayesian MMM for marketing attribution and optimization in a data-driven company like Lemonade. The approach is flexible, interpretable, and can provide valuable insights for decision-making.

## 1. Introduction

In today's competitive business landscape, insurance companies face unique challenges when it comes to effectively allocating their marketing resources. As privacy becomes a driving force in the digital landscape, and user tracking is becoming less accurate [3], it has become imperative for insurance companies to adopt advanced analytical approaches to optimize their marketing strategies without user tracking. In this paper, we propose the development of a "Bayesian Marketing Mix Modeling" model tailored specifically for Lemonade, an insurance company known for its innovative approach to the industry with cutting-edge technology and a seamless digital experience to their customers.

Marketing mix modeling (MMM) has long been recognized as a powerful tool to assess the impact of various marketing channels on business outcomes [4]. By quantifying the effectiveness of each marketing component, MMM helps businesses make informed decisions about resource allocation and campaign optimization.

In this paper, we present a two-step Bayesian MMM approach that benefits from both carryover and shape effects [1], as well as time-varying coefficients [5] that allow the model to recognize trends in marketing channels as well as whole market trends, and model non-linear effects of monetary spending.

Our primary objective is to leverage MMM to enhance Lemonade's marketing effectiveness by providing actionable insights into channel performance, budget allocation, and customer acquisition strategies. By considering factors such as advertising expenditure, online campaigns, social media engagement, and traditional marketing efforts, the MMM model aims to provide a comprehensive understanding of the marketing mix and its impact on key performance indicators such as customer acquisition and customer lifetime value.

Through the implementation of MMM, Lemonade can make data-driven decisions to optimize their marketing mix, enhance customer targeting, and allocate resources more efficiently.

The remainder of this paper is structured as follows: Section 2 provides an overview of the problem, related literature, and existing approaches to marketing mix modeling. Section 3 outlines the methodology and framework for developing the MMM model. Section 4 presents the data collection and analysis process, while Section 5 discusses the results and implications for Lemonade while summarizing the findings, highlighting the practical implications, and suggesting future directions for research.

# 2. Problem Formulation

## 2.0 Definitions

Throughout the paper, we will be using the following mathematical annotations:

$P$ – the number of marketing channels (regressors).

$T$ – number of days (samples) in the data.

$y_t$ – Our target variable at time $t$ (policies or life-time value[1], in our case).

$x_{t,p}$ – the spend in the marketing channel $p \in [p]$ at time $t \in [T]$.

## 2.1 Output of the Model

The output of the model is **two-fold**.

The **first** is the predicted performance for a given budget. Using this output, simulations of different budgets can be run and evaluated, in search of the best budget plan. Optimization tools can be used for this search and will be discussed in section 3.5.

The second is the "channel contribution" ($E_{t,p}$). This output should be the partial accountability for the performance for each channel at some time $t$. In other words, this output would complete the sentence "Regressor $p$ accounted for $X\%$ of the total performance at day $t$". This knowledge would allow us to evaluate our partnerships with advertising channels and could lead to business decisions regarding which partners to enhance and which to part with. We will go into more depth regarding this definition is section 3.4.

Additional outputs can be generated such as reach and carryover curves (As defined in sections 2.3.1, 2.3.2 respectively) which could give more insight into the advertisement channel behavior; however, these are not the focus of our initial research.

## 2.2 Regression for Time-Series

Marketing Mix Modeling is, in its core, regression for a time-series problem, meaning that it is of the form (6):

$$\hat{y}_t = \sum_{p=1}^{P} \beta_p f(x_{t,p}) + \epsilon_t \qquad Eq.1$$

Advancements in time-series analysis (7) proposed the following decomposition:

$$\hat{y}_t = \sum_{p=1}^{P} \beta_p f(x_{t,p}) + g(t) + s(t) + \varepsilon_t \qquad Eq.2$$

Where $g(t)$ is the **trend** function which models non-periodic changes in the value (in our case, it could be the insurance market trend), $s(t)$ represents periodic changes (weekly/yearly seasonality).

The trend and seasonality components can be extracted using several methods, the most notable of which is called STL (8).

## 2.3 Advertising Effects

In addition to the model being a time-series regression model, what makes an MMM model unique is the addition of two non-linear functions that are applied to our budget. These two functions are called **reach** and **carryover** (1).

---

[1] https://en.wikipedia.org/wiki/Customer_lifetime_value

### 2.3.1 Reach

A common assumption in advertising is that it usually has a "diminishing returns"[2].

As more money is invested in a marketing channel, its effect diminishes as the advertising becomes saturated and the relevant customers are reached.

Reach is a function used to model the diminishing returns on ad spend.

Unlike the paper by google (1) we chose a simpler function that has a single parameter:

$$f_{Reach}(x_{t,p}) = \frac{1 - e^{-\frac{x_{t,p}}{\mu_p}}}{1 + e^{-\frac{x_{t,p}}{\mu_p}}} \qquad Eq.3$$

The effect is visualized in Figure 1.

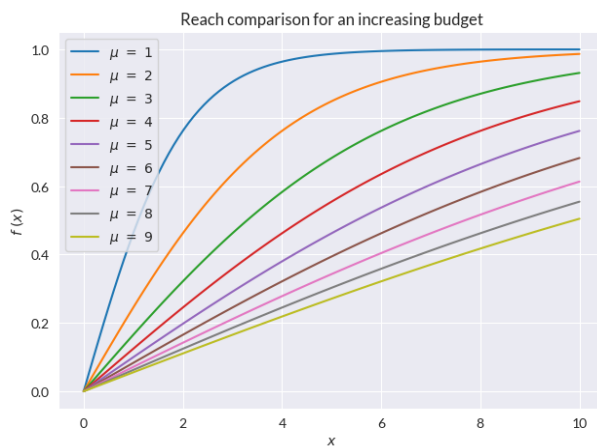

*Figure 1: The Reach function with different parameters u*

This function, applied to our spend data, adds a new parameter to the model that we need to learn – "$\mu_p$", the diminishing returns rate of each regressor.

### 2.3.2 Carryover

The second function - carryover, is used to model the lag effect of advertising.

Some marketing channels, such as television advertising, have delayed effects. For example, a customer could see the television ad during the week, but order insurance at the weekend when the person has time.

To model this effect, we use the carryover (geometric decay) function:

---
[2] https://en.wikipedia.org/wiki/Diminishing_returns

$$f_{Carryover}(x_{t,p}) = \frac{\sum_{l=0}^{t-1} \alpha_p^l x_{t-l,p}}{\sum_{l=0}^{t-1} \alpha_p^l} \qquad Eq.4$$

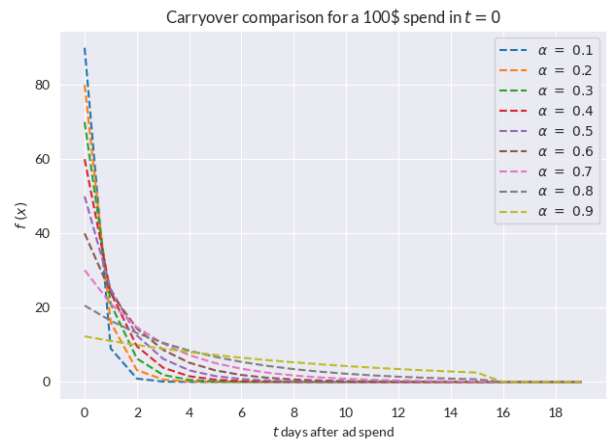

*Figure 2: The Carryover function with different parameters a*

This function also adds a new parameter to the model, $\alpha_p$, the rate of decay for each regressor.

It, is visualized in Figure 2.

### 2.3.3 Combining Both Transformations

Choosing the order in which to apply reach and carryover is explained in the original Google paper (1), and we chose to follow that logic and apply the carryover transformation first since our spend is relatively spread across long time spans.

## 2.4 Time Based coefficients

Another advancement to the basic MMM model is the conversion of the $\beta_p$ coefficient from to equation Eq.2 to a time based $\beta_{t,p}$ (5) that will be impacted by time. This method will be expanded upon in section 2.8.4.

## 2.5 Scaling

As our final output will be the coefficients themselves, scaling the spend $X \to \tilde{X}$ and target $y \to \tilde{y}$ to the same value range will give the coefficients a better relational interpretation between them, and convenience when calculating the prior values and learning the non-linear parameters (1).

A common scaling method which we will use is Max Absolute Scaler:

$$MaxAbsScaler(v) = \frac{v}{\max|v_i|} \qquad Eq.5$$

## 2.6 The Marketing Mix Model

Putting everything together, an MMM model is of the form:

$$\hat{y}_t = \sum_{p=1}^{P} \beta_{t,p} f^*(\tilde{x}_{t,p}, \mu_p, \alpha_p) + g(t) + s(t) \qquad Eq.6$$

Where $f^*(x) = f_{Reach}(f_{Carryover}(x, \alpha), \mu)$

## 2.7 Evaluation

Evaluating an MMM model is difficult. The evaluation can be split between the two outputs of the model:

### 2.7.1 Target prediction evaluation

For evaluating the performance of the target prediction, $R^2$[3] is the preferred and standard metric, as it is a simple metric that captures the time-series relationship between the actual and predicted target.

Additional evaluation metrics we will consider are:

*Mean Absolute Scaled Error*[4]

*Mean Absolute Percentage Error*[5]

We'll also calculate each metric when performing sliding-window cross validation (2).

### 2.7.2 Evaluation with A/B testing

Media Mix Modeling and incrementality testing are interlinked (9). Since the main output of the model is the "contribution" of each marketing channel, or the percentage of the target achieved by the channel, performing an A/B test (for example, shutting down the marketing channel in a subset of geographies (10)) and measuring the incremental change in performance gives us an approximate contribution to compare to.

In this manner, Media Mix Modeling is a predictor for the performance of A/B tests, which are expensive to perform. Thus, the measurements gained by historical A/B tests can be directly applied as a "ground truth" for the calibration of the model and incorporated in evaluation metrics.

## 2.8 Related Work

### 2.8.1 Log-Log frequentist regression

Original MMM models (11) attempted to use the log-log transformation for regular regression to model the diminishing returns as follows:

$$\log \hat{y}_t = \sum_{i=1}^{n} \beta_i \log(x_{i,t}) + \epsilon_t \Rightarrow \qquad Eq.7$$

$$\hat{y}_t = \prod_{i=1}^{n} x_{i,t}^{\beta_i} + \epsilon_t \qquad Eq.8$$

This structure allows for our regression coefficients to also account for the decay in ad effectiveness as a function of spend (as $\beta_i < 1$, since the data is scaled, see chapter 2.5).

Solving for this equation could be done by utilizing Dynamic Linear Models (12) or Kalman filter (13).

### 2.8.2 Bayesian Methods

As computation became more powerful and Bayesian inference became more common, Bayesian modeling became a preferred method (4). Bayesian modeling introduced the ability to model with priors that could **insert prior belief**. In addition, Bayesian modeling generates actual variable **variance** (from the empirical posterior distribution) and can produce a confidence interval to signify which parameters were learned better/worse.

### 2.8.3 PyMC-Marketing

#### 2.8.3.1 Background

A novel and easy to implement approach, PyMC released an open-source Python library called "PyMC Marketing"[6] which contains an out-of-the-box implementation of an MMM model.

#### 2.8.3.2 Method

The model structure can be seen in Figure 3.

The model is built using PyMC and fitted using MCMC methods and (as of the writing of this paper, *v0.2.1*) the No-U-Turn sampler (14).

#### 2.8.3.3 Advantages

The implementation is straight-forward and simple to understand and use; it is easy to tune and adjust the priors.

---

[3] https://en.wikipedia.org/wiki/Coefficient_of_determination
[4] https://en.wikipedia.org/wiki/Mean_absolute_scaled_error
[5] https://en.wikipedia.org/wiki/Mean_absolute_percentage_error
[6] https://www.pymc-marketing.io/

### 2.8.3.4 Disadvantages

The prior distribution cannot be modified, and the regression coefficients aren't time based.

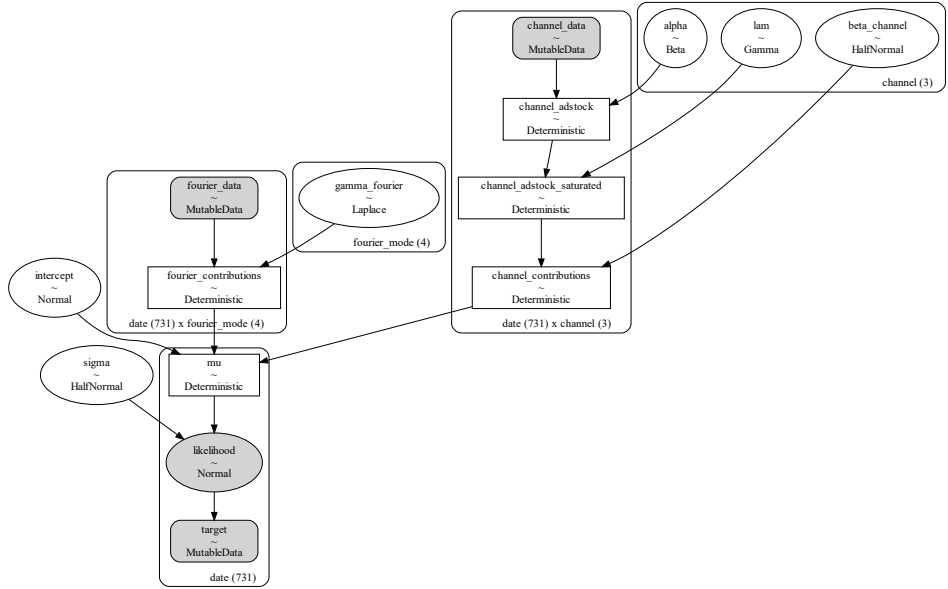

*Figure 3: The general MMM model structure implemented by the library PyMc-Marketing (v0.2.1). In this paper we use the convention mu instead of lam*

It takes a lot of modifications to allow for different sampling algorithms or to use variational inference.

### 2.8.4 Kernel-Based Time-Varying Regression

#### 2.8.4.1 Background

The KTR model, introduced in (5), is built using Pyro and introduces a kernel-based method to calculate time based coefficients ($\beta_{t,p}$).

To calculate the time-based coefficients, the algorithm uses spline regression (15), by sampling the knot values ($b_{j,p}$, $j \in [J]$) and a kernel function to calculate:

$$B_{T \times P} = K_{T \times J} b_{J \times P}$$

Where $B$ is the coefficient matrix $\beta_{t,p} = [B]_{t,p}$.

#### 2.8.4.2 Advantages

Firstly, the channel behavior over-time is learned. Additionally, the model is implemented in Pyro and uses ADVI (Automatic Differentiation Variational Inference (16)) to learn and so converges much faster than MCMC methods.

#### 2.8.4.3 Disadvantages

The model does not take carryover into account and requires a log-log transformation for learning reach, which converts the equation to a multiplicative model (Eq.12) instead of an additive one which makes learning the channel contribution more complex as we cannot just calculate each channel individually.

$$\sum_{p=1}^{P} \beta_p \log x_{t,p} = \log y_t \Rightarrow \prod_{p=1}^{P} x_{t,p}^{\beta_p} = y_t \quad \text{Eq.9}$$

ADVI assumes approximations (such as mean-field (17)) and uses the ELBO loss (18) calculation, which is not necessarily convex. These assumptions entail that the model has less expressive possible posterior distributions.

## 3. Materials and methods

### 3.1 Initial attempts

#### 3.1.1 Robyn and an Internal PyMC mode

We started our research by attempting to use Meta's model "Robyn"[7], but this model did not meet our requirements. We then tried constructing a relatively simple MMM model without scaling while taking inspiration from the implementation of a company called Hello-Fresh (19).

This model gave poor results (both on evaluation metrics such as $R^2$ and on improbable channel contributions that do not align with A/B tests) on the daily level but managed to meet our expectations and align with AB-tests when aggregating our data to a weekly level.

---

[7] https://facebookexperimental.github.io/Robyn/

A further enhancement was attempted to add time-varying coefficients by defining the coefficient as a $\beta_{\cdot,p} \sim GaussianRandomWalk$ however this model gave poor results as well.

### 3.1.2 KTR model implementation

After gaining a baseline PyMC model that works fine for weekly data we set forth on adding what we consider as mandatory features – time-varying coefficients on daily data.

The first version of the model on daily data with the log-log transformation showed a lot of promise, but we ran into complications when trying to extract the contribution of the channels.

We attempted different manipulations of the data to extract the partial contribution of each channel, including attempting to calculate the SHAP values for multiplicative models using the equations developed by Bouneder L. and Leo Y. (20) however all the methods we attempted gave unsatisfactory results.

### 3.2 Our Proposed Solution

After realizing that the log-log transformation denies us one of the two crucial goals of the project, we wanted to use max absolute scaling instead. However, using this method with the KTR mode; means that we don't model carryover or reach. To overcome this issue, we decided to "stack" both models together, meaning that we perform the following modeling flow:

$$x \xrightarrow[PyMC]{\substack{carryover \\ reach}} f(x) \xrightarrow[KTR]{\substack{time-varying \\ coefficient}} \beta_{t,p} f(\tilde{x}) \qquad Eq.10$$

To realize this methodology, we needed to convert our PyMC model to daily data, and for that we replaced it with the newly developed (as of the writing of this paper) PyMC-Marketing package and the model "Delayed Saturated MMM".

This worked well and allowed us to learn the carryover ($\alpha$) and reach ($\mu$) parameters using a MCMC based algorithm, and then learn the time-varying coefficients using variational inference and a kernel-based method which gives us a model that "ticks all the boxes".

### 3.3 Priors

For the regression coefficient, we make the following assumption:

$$\beta_{t,p} \sim \mathcal{N}^+(\beta_{t-1,p}, \sigma_p)$$

Using max absolute scaling, which is a linear transformation, made calculating the prior quite simple and we can infer it as follows:

$$\beta_{0,p} = \frac{\overbrace{P}^{\substack{half-normal \\ scaling}}}{\sqrt{1-\frac{2}{\pi}}} \cdot \overbrace{\frac{\max|y_{\cdot}|}{\max|x_{\cdot,p}|}}^{\substack{data \\ scaling}} \cdot \frac{1}{T}\sum_{t=1}^{T} \frac{x_{t,p} \cdot y_t}{\sum_{p=1}^{P} x_{t,p}} \qquad Eq.11$$

Using the assumption that our historical spend share should be approximately correct (a channel that gets 20% of the budget is expected to have a higher contribution than a channel with 3% of the budget), This calculation takes the average (scaled) spend share multiplied by the performance.

To get the prior standard deviation of $\sigma_p$, we calculated the sample variance instead of the average in the same manner.

For the priors of $\alpha, \mu$ we predefined a prior for each segment of the marketing funnel, a topic which is out of the scope of the paper but in summary we have prior knowledge of how advertising channels behave over time (for example television ads which are higher in the marketing funnel are expected to have a larger carryover effect when compared to Google search ads which are at the bottom of the funnel).

### 3.4 Output of the model

#### 3.4.1 Channel Contribution

Thanks to the linear scaling and an additive model, the contribution of each channel $p$ at any time $t$ can be derived as specified below.

Sample $N$ times from the learned posterior distribution:

$$\hat{\beta}_{tpn} \sim P(\hat{\beta}_{t,p}|X) \quad \hat{\alpha}_{pn} \sim P(\hat{\alpha}_p|X), \hat{\mu}_{pn} \sim P(\hat{\mu}_p|X), \forall n \in [N]$$

We'll define the channel contribution at time $t$ for channel $p$ and sample $n$ as:

$$E_{tpn} = \hat{\beta}_{tpn} \cdot f^*\left(\frac{x_{t,p}}{\max|x_{\cdot,p}|}, \hat{\mu}_{pn}, \hat{\alpha}_{pn}\right) \cdot \max|y_{\cdot}| \qquad Eq.12$$

Calculate the expected value of the channel contribution at time $t$ for channel $p$:

$$\mathbb{E}[E_{t,p}] = \frac{1}{N}\sum_{n=1}^{N} E_{tpn} \quad Eq.13$$

And the standard deviation of the channel contribution:

$$\sigma[E_{t,p}] = \frac{1}{N-1}\sum_{n=1}^{N}(E_{tpn} - \mathbb{E}[E_{tpn}])^2 \quad Eq.14$$

### 3.4.2 Forecasting

Forecasting in the model is currently done in a trivial fashion, where we are given a date range (future or past) and a budget B, we assume that the budget is evenly distributed across the date range and predict using Eq.6.

## 3.5 Optimization methods

Using this forecasting capability, we can attempt to find the best budget allocation by solving the following optimization problem (1):

$$\underset{x}{maximize} \sum_{t=T^S}^{T^E}\sum_{p=1}^{P} \beta_{t,p} f^*(x_{t,p})$$
$$s.t.$$
$$x \in \mathbb{R}^{(T^E-T^S)\times P} \quad Eq.15$$
$$\sum_{t=T^S}^{T^E}\sum_{p=1}^{P} x_{t,p} = B$$
$$T^S < T^E$$

Where $T^S, T^E \in [T]$

### 3.5.1 Knapsack

Initially we saw the optimization as reducible to the knapsack algorithm, where we are looking to "collect" as much value as we can.

This method worked but had drawbacks, we needed to define a step size (minimal budget to move at each iteration) and could potentially have to search the whole possible grid of solutions and recommend a budget allocation that is not business feasible and too different from our original budget.

### 3.5.2 SLSQP

Our second iteration of work on the optimization algorithm led us, along with technical run time considerations, to the "sequential least squares quadratic programming" (21) (SLSQP) algorithm.

This algorithm gave us the option to add restrictions to our possible solutions (for example, we could limit the optimization to only consider budget allocations that deviate up to 20% from the original budget) which helped our growth team gain insights that are actionable and implement changes that are not too drastic.

# 4. Results

## 4.1 The data

In this paper we'll use synthetic data, produced using the code in the corresponding repository in GitHub[8].

The data consists of samples across two and a half years of marketing spend, by week, in 2 channels. The first channel behaves similarly to a long-carryover channel such as tv, and the second channel behaves like search-engine advertising (short-carryover).

To simulate the data, we chose the following priors:

|       | Saturation ($\mu$) | Carryover ($\alpha$) | Regression Coefficient ($\beta$) |
|-------|---------|----------|------------------------|
| $X_1$ | 4.0     | 0.4      | 3.0                    |
| $X_2$ | 3.0     | 0.2      | 2.0                    |

Using these priors, we generated the following channel data seen in **Figure 5**.

In addition, we added seasonality, trend and some noise and added all the components to generate our target (**Figure 4**).

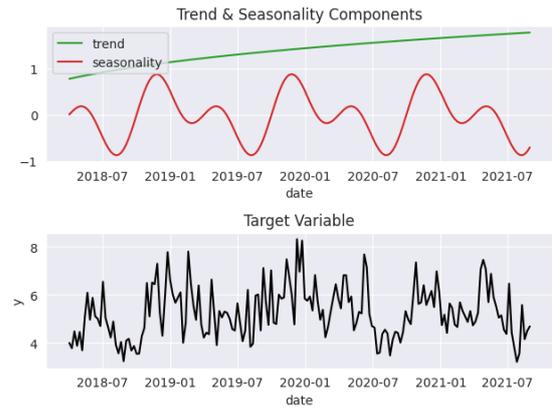

Figure 4: Trend, seasonality components and target

---

[8] https://github.com/royrvd/final-project-in-data-science

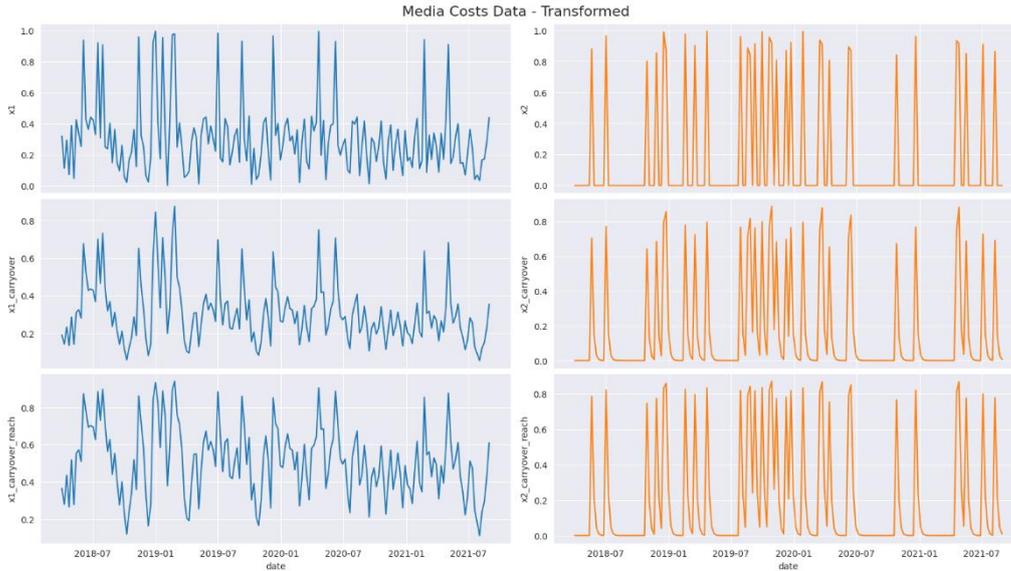

*Figure 5: Channel Data before and after transformation*

## 4.2 First Layer: PyMC-Marketing

After generating the data, we train a PyMC-Model[9]. This is the most time consuming step as MCMC methods take a relatively long time to run.

## 4.3 Second Layer: KTR

Using the posterior distribution of the parameters we learned using the first layer, we transform our data using carryover and reach and train the KTR model on the transformed data.

## 4.4 Model Results

The model results are measured in two aspects, as discussed in 2.1.

### 4.4.1 Forecast Metrics

On a test set of 30 weeks, we got a score of:

$R^2 = 0.47, \quad MAPE = 0.11, \quad MASE = 0.59$

And the fit can be seen in **Figure 6.**

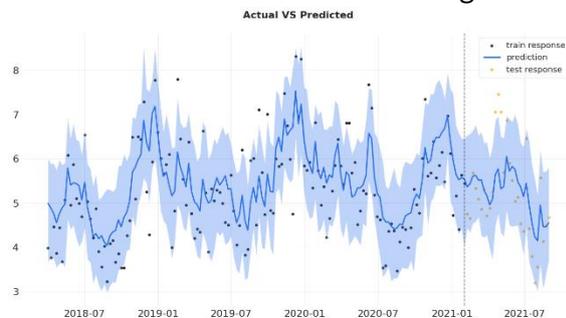

*Figure 6: Model Prediction vs Actual*

### 4.4.2 Channel Contribution

For our coefficient estimates, the expected posterior is:

|       | $\mathbb{E}[\hat{\mu}]$ | $\mathbb{E}[\hat{\alpha}]$ | $\mathbb{E}[\hat{\beta}]$ |
|-------|------|------|------|
| $X_1$ | 4.66 | 0.29 | 3.22 |
| $X_2$ | 2.86 | 0.32 | 3.81 |

Using these posterior parameters, we can calculate the individual contribution of each channel as explained in section 3.4.

## 4.5 Comparison to Naïve Methods

We can compare the model's performance to each layer individually in Figure 7[10] and in Figure 8 and see that while the general forecast metrics are the worst-performing of the bunch (with possible over-fitting of the PyMC model), its posterior parameter estimation is by far the best (as the KTR model doesn't support reach and carryover).

|        | PyMC-Marketing | KTR   | Our Model |
|--------|----------------|-------|-----------|
| $R^2$  | 0.898 | 0.715 | 0.552 |
| $MAPE$ | 0.056 | 0.094 | 0.116 |
| $MASE$ | 0.281 | 0.449 | 0.593 |

*Figure 7: Evaluation metrics in model comparison*

---

[9] This model and all those following use the default hyper parameters and were not tuned for the synthetic data.

[10] Evaluation metrics were calculated on training data as PyMC-Marketing doesn't support forecasting out of the box as of the time of writing the paper.

| | actual | PyMC-Marketing | KTR | Our Model |
|---|---|---|---|---|
| $\hat{\alpha}_1$ | 0.4 | 0.29 | - | 0.29 |
| $\hat{\alpha}_2$ | 0.2 | 0.32 | - | 0.32 |
| $\hat{\mu}_1$ | 4 | 4.66 | - | 4.66 |
| $\hat{\mu}_2$ | 3 | 2.86 | - | 2.86 |
| $\hat{\beta}_1$ | 3 | 0.32 | 0.54 | 2.72 |
| $\hat{\beta}_2$ | 2 | 0.23 | 1.42 | 4.6 |

*Figure 8: Posterior estimation in model comparison*

# 5. Discussion

## 5.1 Retrospective

The Bayesian Marketing Mix Modeling (MMM) approach employed in this study has yielded valuable insights into the marketing dynamics of Lemonade, shedding light on the effectiveness of various marketing channels and their impact on key performance indicators.

**Channel Contribution**: The channel contribution estimates offer deeper insights into the relative importance of each marketing channel. These estimates provide a clear view of how much each channel contributes to the overall performance of Lemonade's marketing efforts. By understanding the contribution of each channel, Lemonade can make informed decisions about resource allocation. For instance, if the model reveals that social media advertising has a disproportionately high contribution when compared to other channels, the company can consider increasing its investment in this area. This allocation strategy is essential for optimizing budget allocation and ensuring that marketing resources are channeled effectively towards the most impactful strategies.

**Forecast Metrics**: Beyond parameter estimation, the forecast metrics provide a measure of the model's predictive performance. As this is the secondary output of the model, we preferred less predictive models that describe the inner components better. However, this functionality is never-the-less useful and can allow us to forecast our budget performance and even optimize our allocation using the methods specified in section 3.5.

**Comparison to Naïve Methods**: It is worth noting that our Bayesian MMM approach outperforms individual layers when compared to naïve methods. While the general forecast metrics might seem less competitive individually, the combined model offers superior parameter estimation. This underscores the value of a multi-layered Bayesian approach in capturing the complexities of marketing dynamics, including long-term carryover effects and short-term reach effects.

In summary, the results from our Bayesian MMM model empower Lemonade with actionable insights to optimize their marketing strategy. The forecast metrics demonstrate the model's predictive accuracy, while the channel contribution estimates provide a roadmap for resource allocation. This holistic approach reinforces the effectiveness of Bayesian MMM in guiding data-driven decisions for marketing optimization in the dynamic landscape of a digital-first insurance company like Lemonade.

## 5.2 Conclusion

1. Driving Growth: Our study demonstrates that the implementation of marketing mix modeling at Lemonade has been instrumental in understanding the impact of various marketing elements on our business outcomes. This, in turn, has paved the way for more informed and strategic decision-making.
2. Data-Driven Decision-Making: Marketing mix modeling has firmly established the importance of data-driven decision-making within our organization. The results indicate that relying on data and analytics to allocate resources and optimize marketing strategies can lead to improvements in ROI and replace costly AB-tests.
3. Optimizing Marketing Spend: Through marketing mix modeling, we have gained insights into the optimal allocation of our marketing budget across different channels and campaigns. This not only maximizes the efficiency of our marketing spend but also ensures that we are reaching our target audience effectively.

4. Privacy-Centric Approach: Implementing the model gave us another method to measure our marketing, while avoiding using customer level data and remaining in anonymous spending data.
5. Continuous Improvement: Marketing mix modeling is not a one-time endeavor but an ongoing process. As we move forward, our budget should approach an optimal allocation.
6. Cross-Functional Collaboration: The success of marketing mix modeling relies on cross-functional collaboration within the organization. This approach encourages teams from marketing, finance, analytics, and other departments to work together, fostering a culture of data-driven decision-making.

In conclusion, marketing mix modeling at Lemonade has not only provided valuable insights into the impact of marketing efforts but also positioned the company for sustained growth and competitiveness in the market. It is essential to recognize that the success of these efforts hinge on a commitment to data-driven decision-making and a culture of continuous improvement.

## 5.3 Limitations

There are several note-worthy limitations of the proposed model.

Firstly, we make the simplifying (and probably wrong) assumption that individual channel spend does not impact each other. One example of where that assumption is likely to be wrong is in TV channel spend. As this channel raises the awareness of the Lemonade brand, it is likely that exposed users will search for lemonade using search engines with brand keywords such as "Lemonade Insurance", which will increase the spend in brand-based channels. Thus, modeling them separately in an additive model assumes no impact between them which is likely to be false.

Secondly, we do not incorporate small scale and new channels. These channels lack enough data to be modeled well. However, we can mitigate this limitation but occasionally attempting to add the channels and by measuring the models' confidence in them (using the channel contribution standard deviation as explained in section 3.4.1 and Eq.14) before incorporating the channels in the main model.

## 5.4 Looking forward

After developing our initial model, we have our sights set on several major improvements:

- Geo-Level modeling
- Nested hierarchical modeling

### 5.2.1 Geo-Level Modeling

As our model and confidence matures, we can move on to model the marketing mix of sub-national geographies (states in the United States, for example).

We assume that the contribution of channels changes across different sub-geographies and that our current outputs represent the average across all of them. We believe that modeling at a lower geographical level could give more precise and effectful recommendations to our growth teams.

One approach that currently supports such a feature was implemented in the package "Light weight MMM" by Google (22).

### 5.2.2 Nested Hierarchical Modeling

Another known issue we are currently not tackling is the effect that different advertising channels have on each other.

For example, we know that "upper funnel"[11] marketing has a direct effect on organic traffic (non-paid activities) that can be mistaken instead of the channel effect. This is demonstrated by Figure 9.

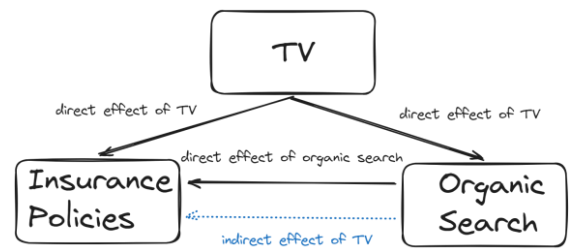

Figure 9: indirect effect of the upper funnel on the lower funnel

One way to tackle this issue is by creating a nested MMM where we also predict the effect of upper-funnel marketing on the activity on the lower funnel and incorporate that as another variable.

---
[11] https://en.wikipedia.org/wiki/Purchase_funnel

This method was suggested to us through conversations with Meta and their Robyn development team.

## 6 Acknowledgements

I would like to thank Adiel Loinger and Uriel Vinetz for their help on this project and Udi Ryba for his business side perspective and input throughout the process.